\documentclass[18pt,a4paper,fleqn,usenatbib,useAMS]{mnras}

\usepackage{newtxtext,newtxmath}

\usepackage[T1]{fontenc}
\usepackage{ae,aecompl}

\DeclareRobustCommand{\VAN}[3]{#2}
\let\VANthebibliography\thebibliography
\def\thebibliography{\DeclareRobustCommand{\VAN}[3]{##3}\VANthebibliography}

\usepackage{graphicx}	
\usepackage{amsmath}	
\usepackage{multicol}   
\usepackage{bm}		    
\usepackage{pdflscape}	
\usepackage{float}
\usepackage{caption}

\usepackage{placeins}
\usepackage{tabularx,booktabs} 

\title[CH$_3$OH]{Physical properties of methanol (CH$_3$OH) ice as a function of temperature: Density, infrared band strengths, and crystallization.} 

\author[H. Carrascosa et al.]{
H. Carrascosa$^{1}$\thanks{E-mail: hcarrascosa@cab.inta-csic.es (HC)}
M. \'{A}. Satorre,$^{2}$
B. Escribano,$^{1}$ 
R. Mart\'in-Dom\'enech,$^{1}$ and
G. M. Mu\~{n}oz Caro$^{1}$
\\
$^{1}$Centro de Astrobiolog\'{\i}a (CSIC-INTA), Ctra. de Ajalvir, km 4, Torrej\'on de Ardoz, 28850 Madrid, Spain\\
$^{2}$Centro de Tecnolog\'{\i}as F\'{\i}sicas, Universitat Polit\`{e}cnica de Val\`{e}ncia, Plaza Ferr\'{a}ndiz-Carbonell, 03801 Alcoy, Spain}

\date{Accepted XXX. Received YYY; in original form ZZZ}

\pubyear{2023}

\begin{document}
\label{firstpage}
\pagerange{\pageref{firstpage}--\pageref{lastpage}}
\maketitle

\begin{abstract}
The presence of methanol among the common ice components in interstellar clouds and protostellar envelopes has been confirmed by the James Webb Space Telescope. Methanol is often detected in the gas phase toward lines of sight shielded from UV radiation.
We measured the volumetric density of methanol ice, grown under simulated interstellar conditions, and the infrared spectroscopy at different deposition temperatures and during the warm-up. The IR band strengths are provided and the experimental spectra are compared to those computed with a model. The transition from amorphous to crystalline methanol ice was also explored. Finally, we propose new observations of methanol ice at high resolution to probe the methanol ice structure. 
\end{abstract}



\section{Introduction} \label{sect:intro}

Infrared observations allow the identification of simple molecular species in ice mantles. Methanol (CH$_3$OH) is one of the abundant components detected toward numerous inter- and circum-stellar sources (e.g., \cite{Chiar1995}; \cite{Dartois1999}; \cite{Pontoppidan2003}; \cite{Pontoppidan2004}, \cite{Bottinelli2010}; \cite{Boogert2015}; \cite{Boogert2022}). The formation of methanol in the ice by hydrogenation of CO has been proposed (e.g., \cite{Tielens1997}; \cite{Watanabe2002}). In this case, methanol might coexist with CO in ice mantles. Warm-up of icy dust grains as they move into warmer regions would lead to partial sublimation of the more volatile components and a relatively rich methanol phase would remain. The CO hydrogenation hypothesis as the main source of methanol has been questioned (e.g. \cite{Dartois1999}), suggesting that methanol might also form in irradiated ice mixtures as in laboratory experiments, while an efficient gas phase synthesis seems very unlikely. Recent observations also suggest that hydrogenation of CO cannot account for the methanol observed in ice mantles toward dense clouds (e.g. \cite{McClure2023}). There is a fair number of publications devoted to the simulation of methanol ice processes in the laboratory. Energetic processing of pure methanol ice, or methanol-bearing ice mixtures, using X-ray and UV photons (\cite{Allamandola1988}, \cite{Bernstein1995}, \cite{MunozCaro2003}, \cite{Oberg2009}, \cite{Chen2013}, \cite{Basalgete2021}) or ions  (\cite{Palumbo1999}, \cite{MunozCaro2014}) showed that a plethora of complex organic species is formed. The UV photon-induced desorption of methanol ice was explored (\cite{CruzDiaz2016}, \cite{Bertin2016}). 
Simulated cosmic-ray sputtering seems to be the most efficient way to desorb methanol molecules from the ice (\cite{Dartois2019}, \cite{Dartois2020}).  
Other experimental works were dedicated to mimic the thermal desorption of pure methanol ice in astrophysical environments (\cite{Sandford_Allamandola1993}; \cite{Collings2004}; \cite{Green2009}), methanol and water in the ice (\cite{Wolff2007}; \cite{Bahr2008})
or mixed in the ice with CO$_2$ (\cite{Mate2009}) and methanol mixed with other species (\cite{MartinDomenech2014}).\\ 

In this paper, we reproduce experimentally the accretion of methanol ice under ultra-high vacuum conditions. The volumetric density of methanol ice was measured at different temperatures during warm-up. A previous work reported the density of methanol ice samples that were deposited at different temperatures and their optical constants \citep{Luna2018}. The infrared spectroscopy of methanol ice samples at different deposition temperatures and during the warm-up is reported. The IR band strengths are revisited and the experimental spectra are compared to those computed with a model to assess their vibrational modes. Special attention is paid to the previously observed transition from amorphous to crystalline methanol ice and its effect on the ice density and spectroscopy. New observations of methanol ice at high resolution with JWST are proposed to probe the methanol ice structure based on the splitting of infrared absorption bands detected in our work. Meanwhile a more precise determination of these band positions would serve to better constrain the molecular environment of the CH$_3$OH molecules in the ice, which is currently identified as either pure CH$_3$OH or a CH$_3$OH:CO ice mixture. We note that the correct identification of the methanol phase in astrophysical ice mantles is of paramount importance because it shapes the network of chemical reactions triggered by radiation and thermal processing which ultimately lead to complex organic molecules (COMs) of astrobiological significance, see the above-cited references.\\

\section{Experimental} 

Experiments were carried out in ISAC, described in detail in \cite{MunozCaro2010}. ISAC is an ultra-high vacuum (UHV) chamber with a base pressure in the 10$^{-11}$ mbar range. A closed-cycle helium cryostat and a Lakeshore model 331 temperature controller connected to a silicon diode allow to control temperature over a KBr substrate from 10 K to 300 K with an accuracy better than 0.1 K. CH$_3$OH (Panreac Qu\'imica S. A. 99.9 \%) was introduced in the chamber from the gas line. Gas line in ISAC is evacuated with a oil-free pump, under a pressure of $\sim$10$^{-3}$ mbar. Then, methanol vapour is introduced in the gas line until a pressure of few mbar is reached. Methanol vapour is directed towards the substrate through a stainless steel tube with an angle of 45$^\circ$. From this position, both laser interferometry and IR spectra can be recorded during deposition of the ice. Additionally, it has been proven that deposition up to an angle close to 45$^\circ$ has not an effect on the ice structure \citep{GonzalezDiaz2019}. Pressure was constant during deposition, and ices were grown for $\sim$1 h (see Table \ref{Table.experiments}). Laser interferometry is measured continously with a He-Ne red laser (632.8 nm, 5.0 mW, 500:1 polarization), and a silicon photodiode power sensor (model S120C) which are placed with an angle of $\sim$6$^\circ$. A Bruker Vertex 70 with a deuterated triglycine sulfate detector (DGTS) was used to measure Fourier-Transformed Infrared Spectroscopy. IR spectra were acquired with a resolution of 2 cm$^{-1}$ and 128 scans that correspond to 4 min acquisition time. This spectral resolution is often used in experimental simulations, as it provides a good relationship between the acquisition time and the quality of the spectra. IR spectra were taking during deposition and warm-up of the ice samples. Therefore, a better resolution is not useful, as it requires larger acquisition time, and the ice sample will have changed during the acquisition itself.\\

\begin{figure*}
    \centering
    \includegraphics[width=\textwidth]{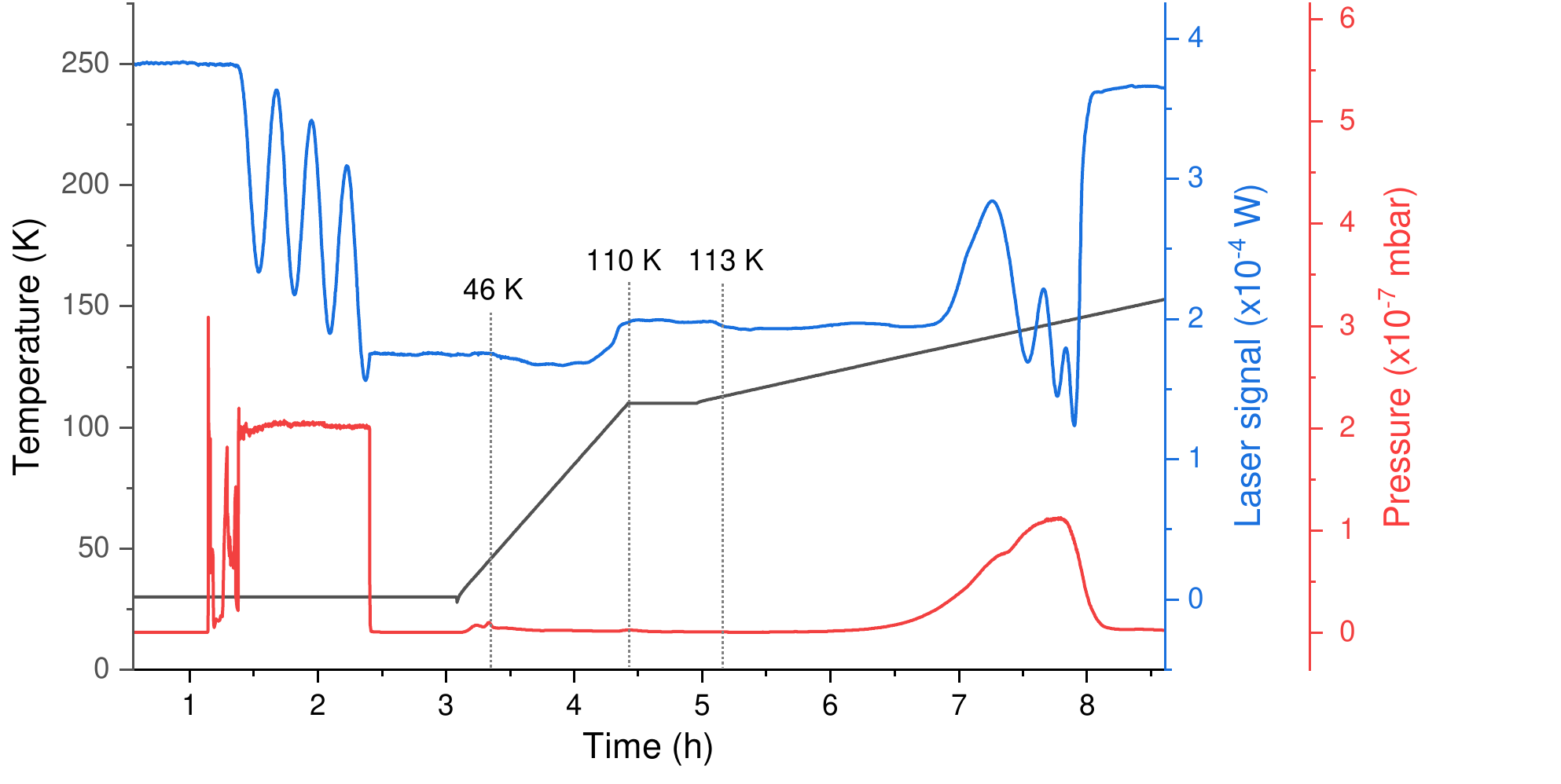}
    \caption{Temperature, laser signal and pressure measured in the chamber during Exp. \textbf{1}. Methanol molecules are responsible for the increase in pressure during deposition of the ice at a constant temperature of 30 K and constant pressure of 2$\times$10$^{-7}$~mbar. Laser interference shows the ice growth during deposition. During warm-up at a rate of 1 K/min (up to 110~K), laser signal shows some variations around 46 and 110~K, related to compactation of the ice (see text). Then the temperature is kept constant during 2000 s to allow stabilization of the ice. From 110 K, warm-up was set at 0.2~K/min until thermal desorption of the ice was achieved.}
    \label{Fig.Resumen_experimento_CH3OH}
\end{figure*}

Pressure inside the chamber is recorded with a Bayard-Alpert sensor, while the composition of the gas phase is also monitored with quadrupole mass spectrometry (QMS, Pfeiffer Vacuum, Prisma QMS 200). Figure \ref{Fig.Resumen_experimento_CH3OH} shows the recorded laser signal, pressure, and temperature during Exp. \textbf{1}.\\

\begin{table} 
    \caption[]{Experimental conditions of the CH$_3$OH ice samples deposited for this work.}
    \label{Table.experiments}
    \resizebox{8.5cm}{!}{
    \begin{tabular}{ccccc}
Exp.         & Temperature (K) & $\Delta$P (mbar) & Time (s) & Warm-up (K/min)\\
\hline
\noalign{\smallskip}
\noalign{\smallskip}
\textbf{1}& 30          & 2$\times$10$^{-7}$            & 3643   & 1  (10-110 K)\\ &&&&0.2  (from 110 K)\\
\noalign{\smallskip}
\textbf{2}& 125          & 2$\times$10$^{-7}$        & 3587     &0.5 \\
\noalign{\smallskip}
\textbf{3}& 135          & 2$\times$10$^{-7}$        & 14200*     &0.1 \\
\noalign{\smallskip}
\hline
\end{tabular}\\
}
\textit{*Note that Exp. \textbf{3} has a larger deposition time, as thermal desorption is already present at that temperature, the overall deposition rate is significantly lower than the one measured for lower temperatures.}

 \end{table}

\subsection{Calculation of band strength and density during warm-up of CH$_3$OH ice}
Density ($\rho$) is the relation between mass and volume. The mass of an ice sample per area unit can be expressed in terms of the molecular mass and the column density, while the volume is related to the thickness of the ice. Therefore, the density obtained at different temperatures $T$ during warm-up can be expressed as follows:

\begin{equation}
   \rho_{CH_3OH}(T) = \frac{m_{CH_3OH} \cdot N(T)}{N_A \cdot d(T)}
   \label{eq.densidad}
\end{equation}\\

{\setlength{\parindent}{0pt}where $m_{CH_3OH}$ is the molecular mass of the ice sample (CH$_3$OH in this work), $N(T)$ is the column density, $N_A$ is Avogadro's number, and $d(T)$ is the thickness of the ice layer. The column density of an ice can be obtained from IR spectrum, considering the band strength ($\mathcal{A}(T)$) of any molecular vibration at a given temperature:}

\begin{equation}
    N (T) = \frac{1}{\mathcal{A (T)}} \cdot \int_{band}\tau_{\nu}(T)d{\nu} = \frac{2.3 \cdot A_{int} (T)}{\mathcal{A} (T)}
    \label{eq.densidad_de_columna}
\end{equation}\\

{\setlength{\parindent}{0pt}where $\tau_{\nu}(T)$ is the optical depth of the band, which is related to the integrated area of the IR band, $A_{int}(T)$. For this work, we have used the most intense IR absorption of methanol ice (centered around 3238 cm$^{-1}$) as it will have the largest signal-to-noise ratio. Substituting \ref{eq.densidad_de_columna} in \ref{eq.densidad} and reorganizing terms:}

\begin{equation}
    \rho_{CH_3OH} (T) = \frac{2.3 \cdot m_{CH_3OH} \cdot A_{int} (T)}{N_A \cdot d (T) \cdot \mathcal{A} (T)}
    \label{eq.density}
\end{equation}

\begin{equation}
    \mathcal{A} (T) = \frac{2.3 \cdot m_{CH_3OH} \cdot A_{int} (T)}{N_A \cdot d (T) \cdot \rho_{CH_3OH} (T)}
     \label{eq.band_strength}
\end{equation}\\

CH$_3$OH ice in Exp. \textbf{1} was deposited at 30 K. A$_{int}(T)$, $m_{CH_3OH}$ and $N_A$ are known, while $d(T)$ and $\rho_{CH_3OH}$ must be determined at 30 K. From the He-Ne laser, $d(T)$ can be obtained \citep[see][for calculation details]{Cristobal_2022MNRAS.517.5744G} assuming a known value of the refractive index of the ice ($n_{CH_3OH}$). To estimate the refractive index and density values at 30 K, the average value between 20 K and 40 K deposition was adopted from \cite{Luna2018AA...617A.116L}, $n_{CH_3OH}$ (30 K) $ = 1.265$ and $\rho_{CH_3OH}$ (30 K) $ = 0.65 $ g cm$^{-3}$.\\

Once the value at the lowest temperature is known, the band strength for any other temperature can be obtained easily, assuming a constant column density during warm-up (Eq. \ref{eq.band_strength_temperatures}). This assumption is thought to be good as long as warm-up does not lead to an increase of pressure inside the chamber, meaning that there is no thermal desorption of the ice. This occurs for temperatures up to 125 K in our experiments.\\

\begin{equation}
    \frac{\mathcal{A} (T)}{A_{int} (T)} = \frac{\mathcal{A} (30~{\rm K})}{A_{int} (30~ {\rm K})} \rightarrow \mathcal{A} (T) = \frac{\mathcal{A} (30~{\rm K}) \cdot A_{int} (T)}{A_{int} (30~{\rm K})}
    \label{eq.band_strength_temperatures}
\end{equation}\\

From the calculated band strength at every temperature, the density of CH$_3$OH for various warm-up temperatures is obtained using \ref{eq.density}.\\

\subsection{Simulation of CH$_3$OH ice structure}
We have carried out theoretical calculations at Density Functional Theory level \citep{DFT-HK,DFT-KS} using the CASTEP software \citep{CASTEP} for geometry optimization and prediction of vibrational spectra. 
The same software was used for generating amorphous ices by annealing the crystalline structure in the NPT ensemble using Nose thermostat \citep{NHC} and Andersen barostat \citep{Andersen1980}.
Geometry optimization was achieved through several stages of progressively higher accuracy, ultimately with energy convergence lower than 1x10$^{-6}$~eV atom$^{-1}$, maximum force of 0.01~eV \AA$^{-1}$ ~and a maximum stress of 0.02~GPa. 
For the DFT calculations we used the Generalized Gradient Approximation (GGA) together with functionals by Perdew-Burke-Ernzerhof (PBE) \citep{Perdew1996}, including Grimme D2 dispersion correction \citep{Grimme} which proved to be very important for overcoming the limitations of PBE when studying hydrogen bonded systems. 
We used the standard norm-conserving pseudopotentials of Troullier Martins \citep{Troullier1991}, as is common when modeling bulk ice, with a plane-wave basis set with an energy cutoff of 830~eV. 
Although PBE functionals have been reported to overestimate the strength of hydrogen bonds in methanol ice \citep{Galvez2009}, we still chose to use it for its universality and applicability.
The BLYP \citep{BLYP} functional was also tested, but for the purpose of this work it was discarded as it produced very similar cell geometry at a higher computational cost.

\section{Results} \label{sect:results}

Fig. \ref{Fig.Resumen_experimento_CH3OH} shows the methanol ice deposition and subsequent warm-up from 30 to 110 K at a rate of 1 K/min. At 110 K the ice temperature is kept constant during 2000 s to allow stabilization of the ice structure. Beyond 110 K, the warm-up rate was set to 0.2 K/min until thermal desorption of the ice was achieved. The three full laser interference cycles (blue trace) measured during methanol ice deposition are also observed during thermal desorption above 125 K, when the molecules in the ice start to desorb and the methanol vapor pressure increases (red trace). The time lapse of each cycle during the deposition is 997 $\pm$ 40 s, which indicates a constant accretion rate of the methanol ice as the result of a constant methanol vapor pressure in the chamber, measured by the Bayard-Alpert gauge (red trace). The duration of the laser cycle (blue trace) decreases during thermal desorption because the desorption rate, in molecules cm$^{-2}$ s$^{-1}$, increases following the exponential curve of the Polanyi-Wigner equation:\\
 
\begin{equation}
\frac{d N_g({\rm CH_3OH})}{d t} = \nu_i [N_s({\rm CH_3OH})]^i \exp(- \frac{E_d({\rm CH_3OH})}{T}) 
\label{polanyi}
\end{equation}

where $N_g({\rm CH_3OH})$ is the column density of CH$_3$OH molecules desorbing from the ice surface $\left(cm^{-2}\right)$, $\nu_i$ a frequency factor $\left(molecules^{1-i} cm^{-2(1-i)} s^{-1}\right)$ for desorption order $i$, $N_s({\rm CH_3OH})$ the column density of CH$_3$OH molecules on the surface at time $t$, $E_d({\rm CH_3OH})$ the binding energy in K, and $T$ the surface temperature in K. The TPD data can be fitted using Eq.~\ref{polanyi} and the relation

\begin{equation}
\frac{d N_g({\rm CH_3OH})}{d t} = \frac{d T}{d t} \frac{d N}{d T} 
\label{temp_vs_time}
\end{equation}
where $\frac{d T}{d t}$ is the warm-up rate provided in Table \ref{Table.experiments} for each experiment.\\

This process is directly observed in the pressure behaviour (red trace). During warm-up, the laser signal (blue trace) shows some variations above 46 K that could be related to a restructuration of the ice. This variation is greater as the ice temperature approaches 100 K and after 113 K.\\

\begin{figure}
    \centering
    \includegraphics[width=0.5\textwidth]{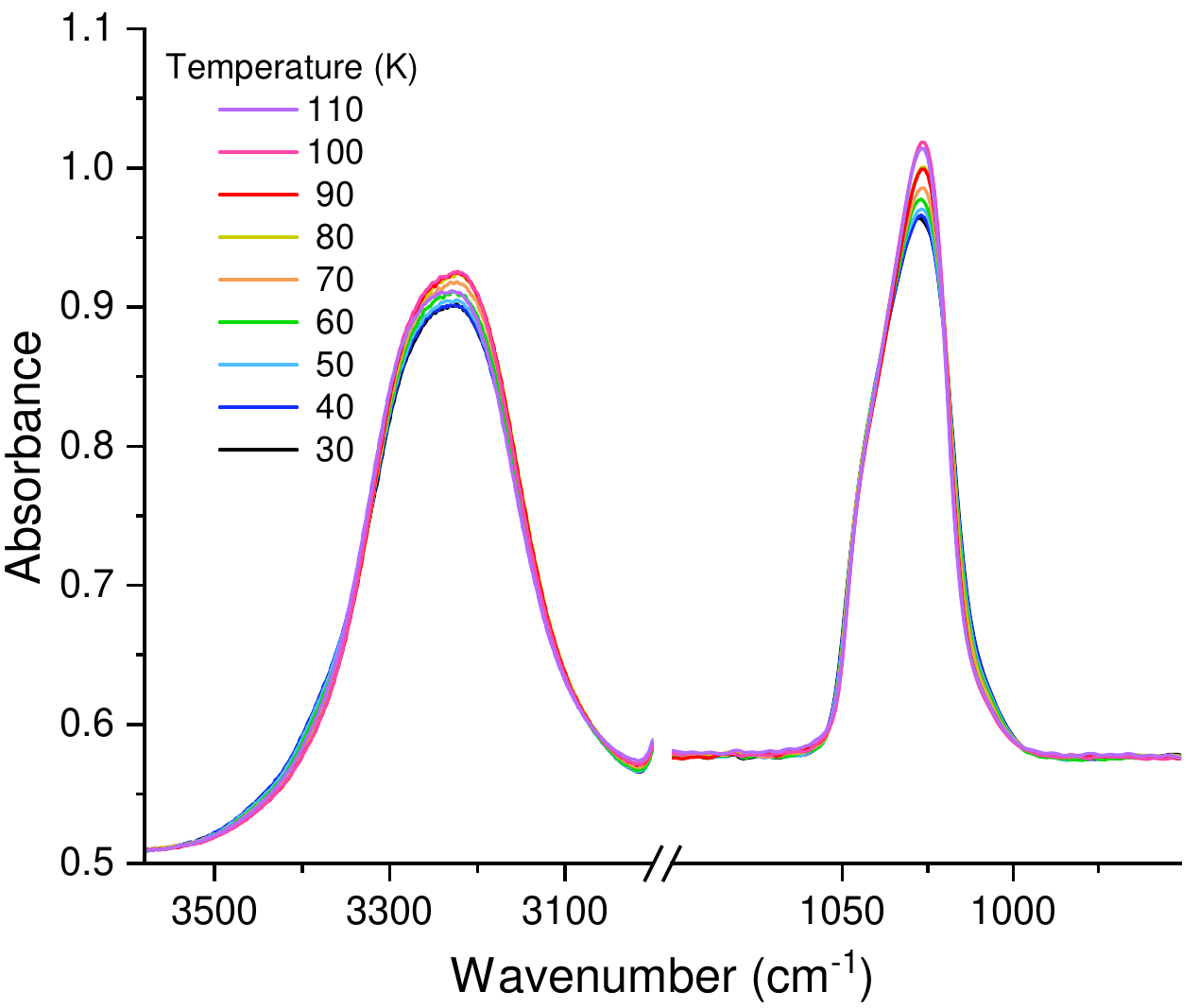}
    \caption{Infrared bands near 3250 cm$^{-1}$, OH stretching, and 1025 cm$^{-1}$, CO stretching, of methanol ice deposited at 30 K during warm-up to 110 K.}
    \label{Fig.Band_intensity}
\end{figure}

Fig. \ref{Fig.Band_intensity} presents the two main infrared absorption bands of 30 K-deposited methanol ice during warm-up to 100 K. These bands are centered near 3250 cm$^{-1}$, OH stretching, and 1025 cm$^{-1}$, CO stretching. Only a gentle increase of the band intensities was observed accompanied by a tiny reduction of the blue/red wing in these bands.\\

\begin{figure*}
    \centering
    \includegraphics[width=0.8\textwidth]{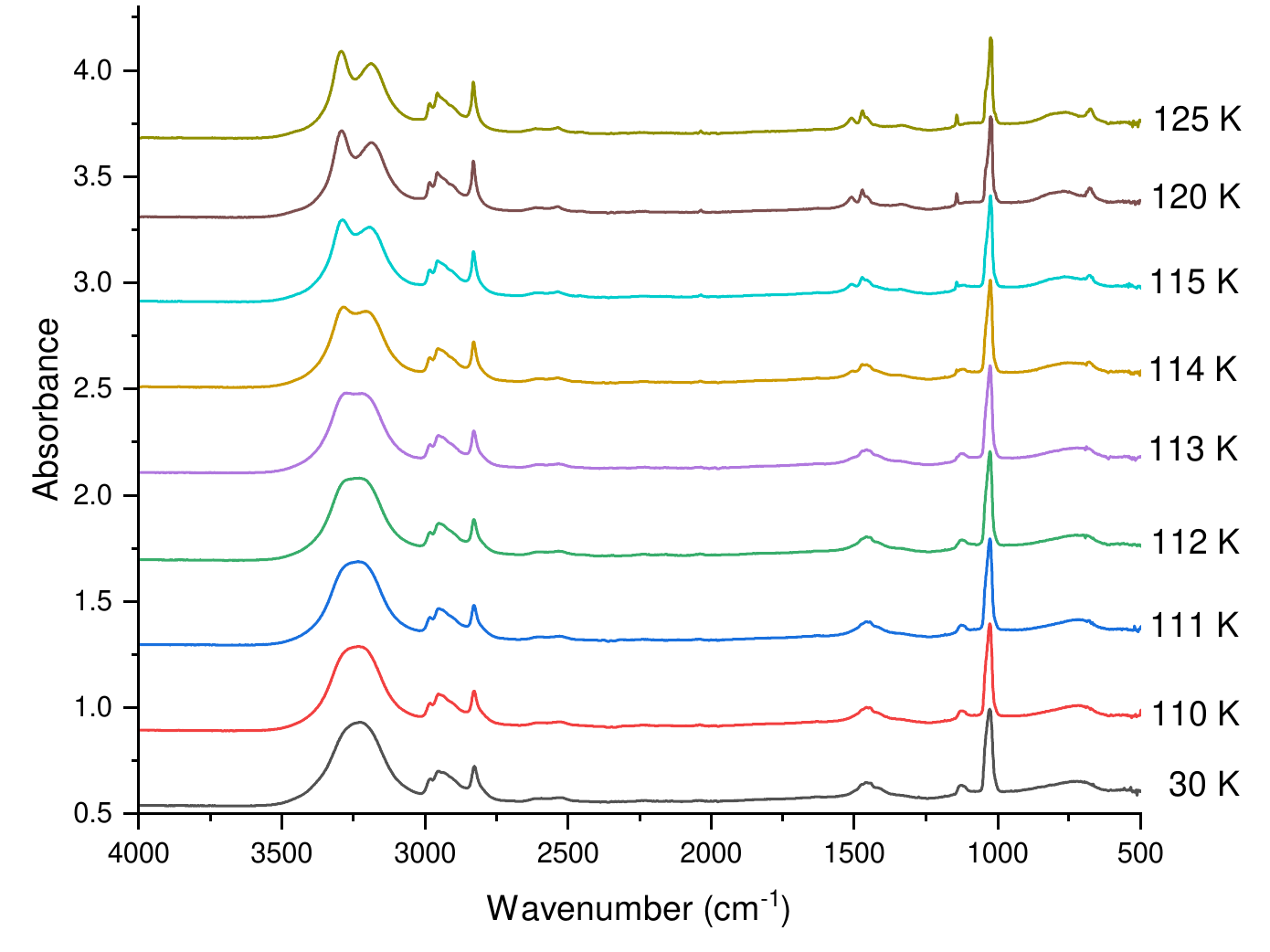}
    \caption{Evolution of the IR spectrum of methanol ice (Exp. \textbf{1}) from deposition at 30 K to crystallization. warm-up from 30 K to 110 K does not substantially change the IR spectra. Above that temperature, a better defined shape is observed, specially for the O-H stretching mode centered at 3238 cm$^{-1}$, which already shows two different contributions from 112-113 K.}
    \label{Fig.IR_cristalizacion}
\end{figure*}

\begin{figure*}
    \centering
    \includegraphics[width=0.8\textwidth]{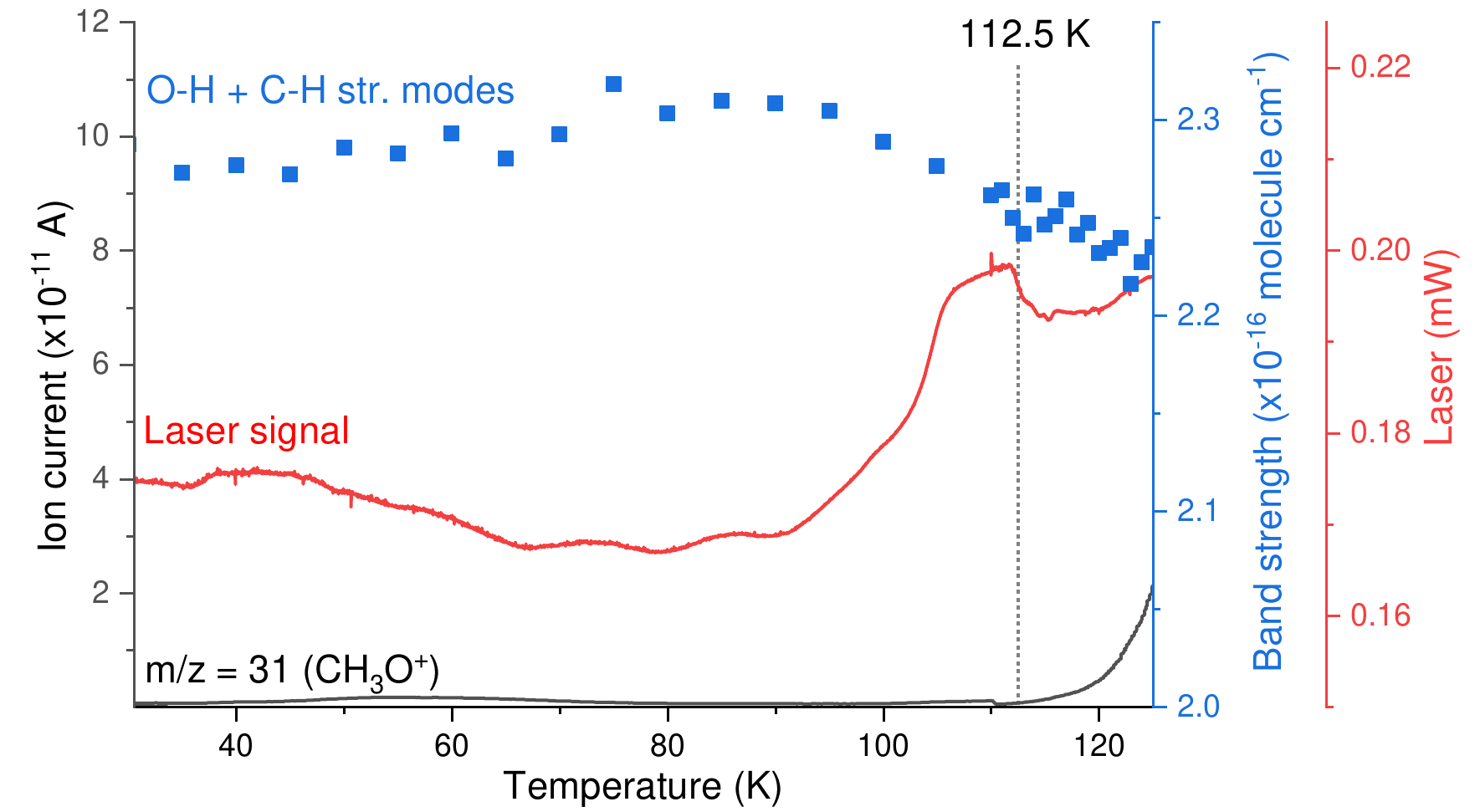}
    \caption{Evolution of the band strength that corresponds to the spectral range covering the O-H and C-H stretching modes of CH$_3$OH ice
    (Exp. \textbf{1}) as a function of the warm-up temperature. Band strength was calculated assuming a constant column density during warm-up. This approximation is valid if thermal desorption of methanol ice is negligible, as it happens up to 125 K (the increment in the ion current from 112.5 K is about 4 orders of magnitude lower than the signal measured during thermal desorption). Before this temperature, there is an important variation in band strength and laser signal around 112.5 K due to the onset of crystallization of the CH$_3$OH ice as derived from IR spectra, shown in Fig. \ref{Fig.IR_cristalizacion}.}
    \label{Fig.Band_strength_CH3OH}
\end{figure*}

\begin{figure*}
    \centering
    \includegraphics[width=\textwidth]{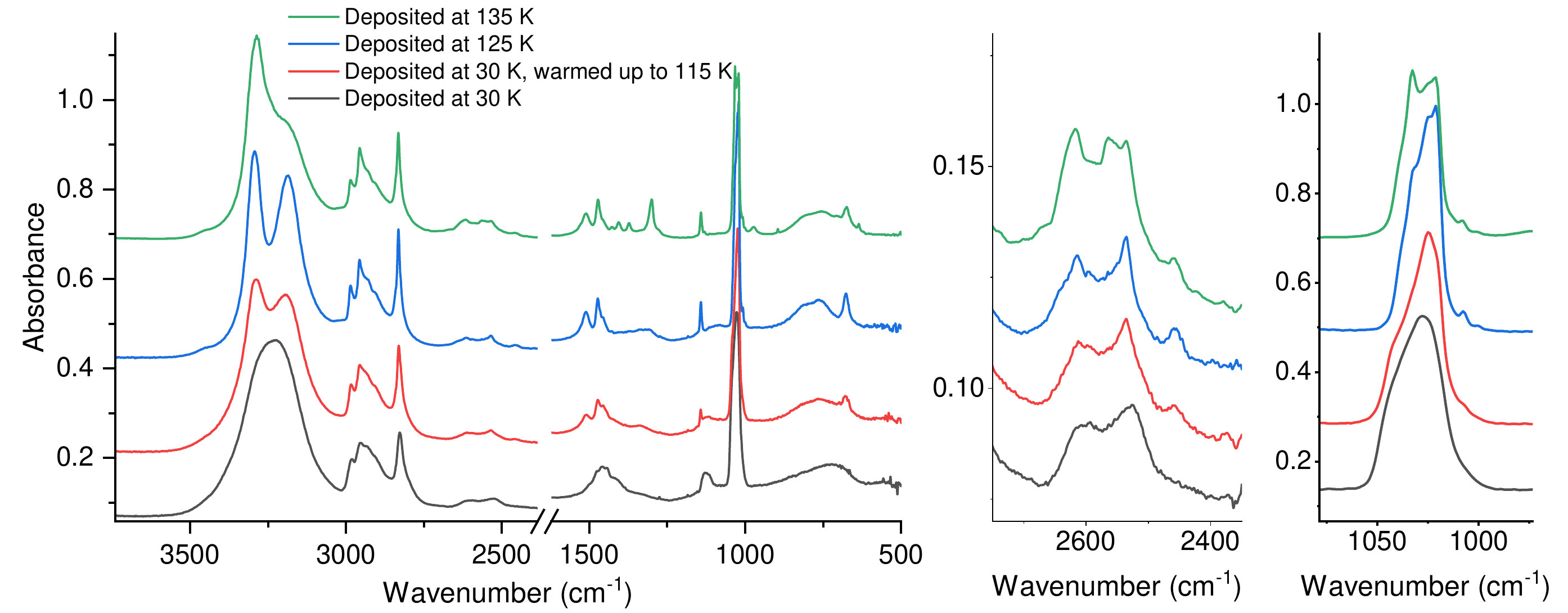}
    \caption{IR spectra recorded for CH$_3$OH ice samples in Exps. \textbf{1} (30 K deposition and 115 K warm-up), \textbf{2} (125 K deposition) and \textbf{3} (135 K deposition). The crystalline structure appearing at 112-113 K during warm-up is more pronounced when deposition of the ice takes place at 125 K. Deposition at 135 K is, however, different, indicative of another crystalline phase for methanol ice, see Sect. \ref{sect:dft}. Middle and right panel show in detail the combination bands between 2700 - 2500 cm$^{-1}$ and the C-O stretching band at 1027 cm$^{-1}$, respectively.}
    \label{Fig.IR_deposition_temperatures}
\end{figure*}

\begin{figure}
    \centering
    \includegraphics[width=0.49\textwidth]{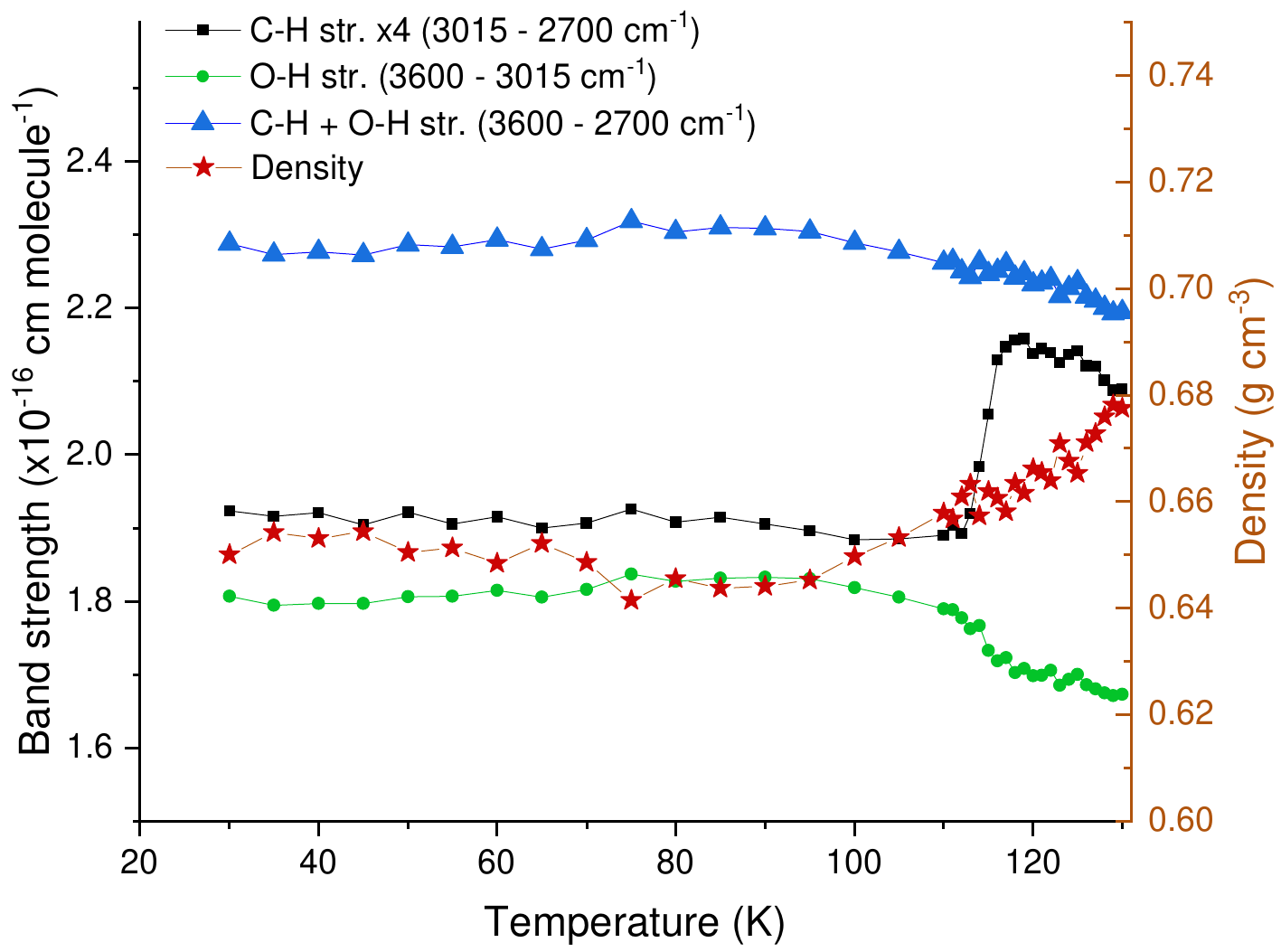}
    \caption{Band strengths for CH str. (black squares, multiplied by 4 for clarity), OH str. (green circles), and the combination of these modes integrating the full spectral region (blue triangles)  at various temperatures during warm-up. The density values for these warm-up temperatures are represented by the red stars.}
    \label{Fig.Band_strength_density}
\end{figure}

\begin{figure}
    \centering
    \includegraphics[width=0.49\textwidth]{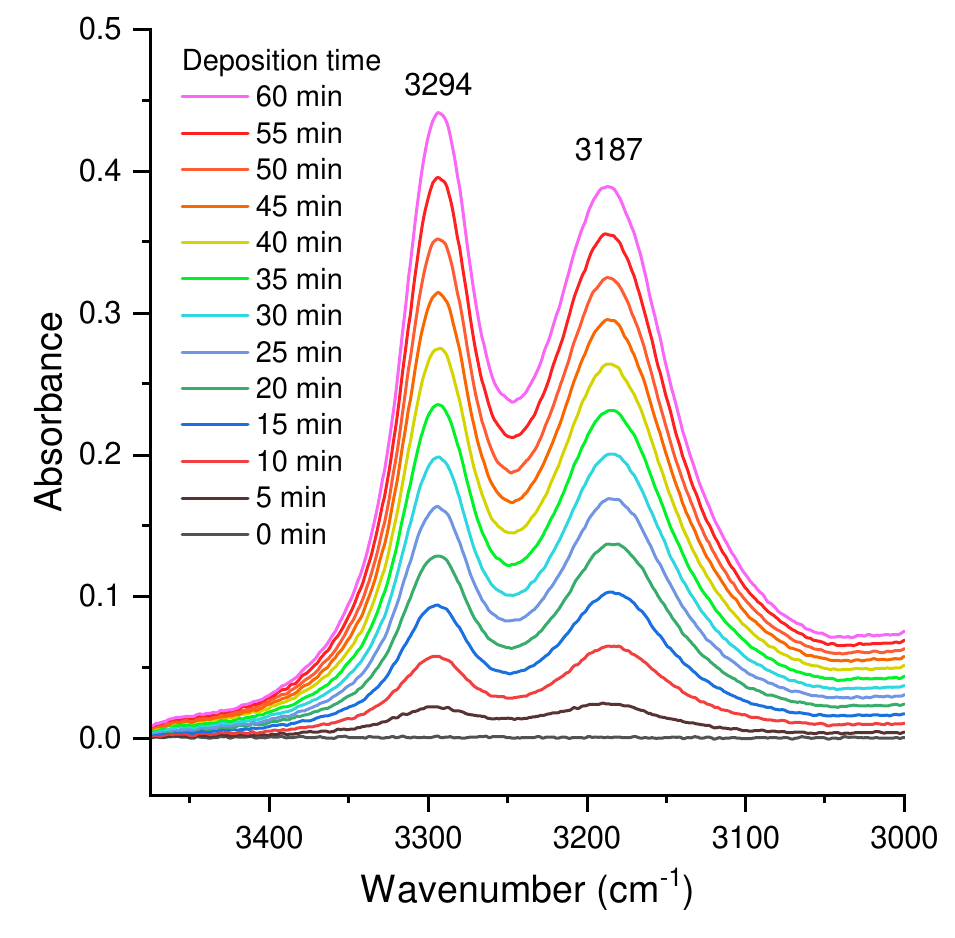}
    \includegraphics[width=0.49\textwidth]{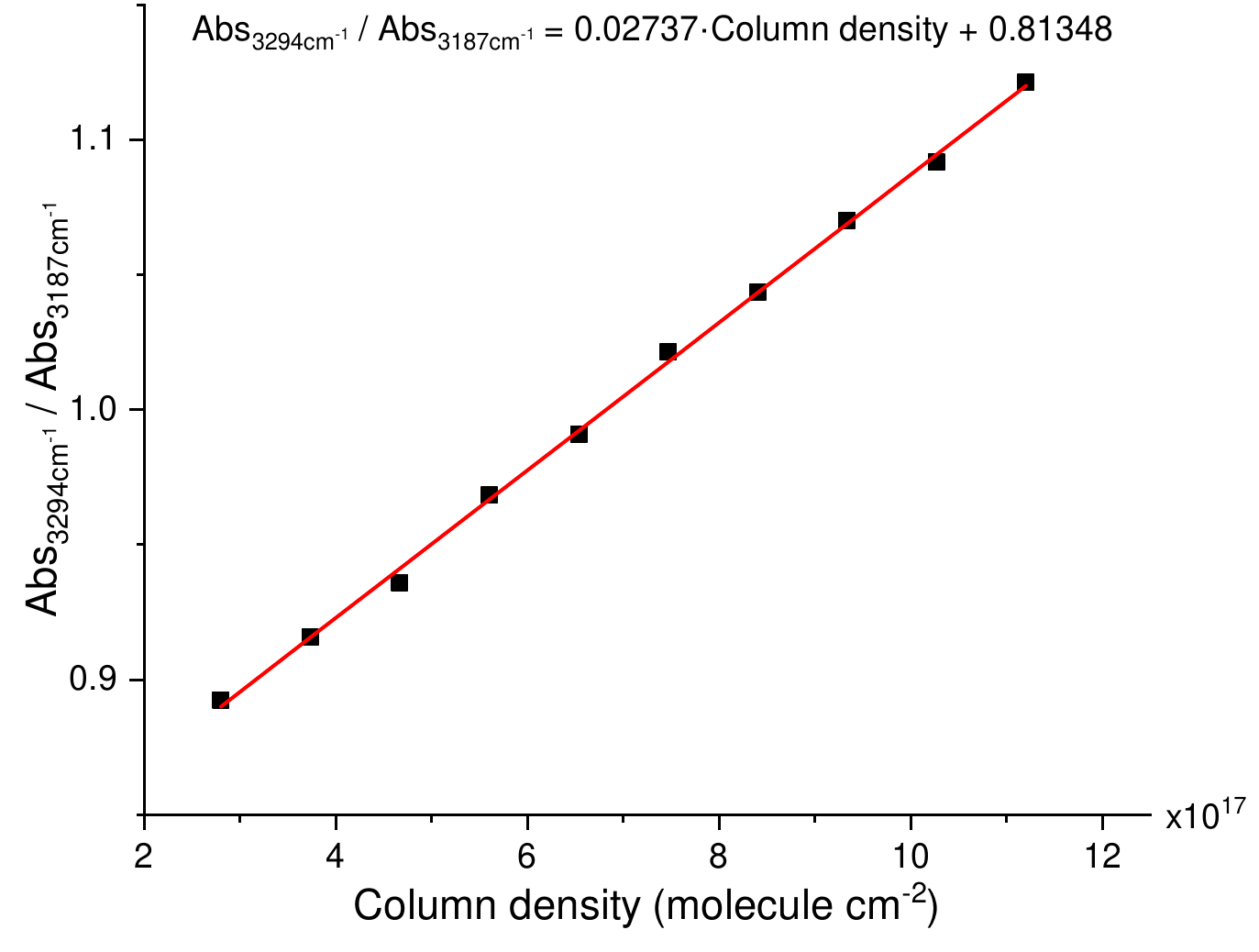}
    \caption{Top panel: IR spectra during deposition at 125 K of a CH$_3$OH ice sample (Exp. \textbf{2}). Along the first steps of deposition, the peak centered at 3187 cm$^{-1}$ is more intense than the one centered at 3294 cm$^{-1}$. As the deposition continues, the 3294 cm$^{-1}$ peak turns out to be more intense than the 3187 cm$^{-1}$ feature. Bottom panel: ratio between the maximum of both peaks as a function of deposition time (expressed as column density) with linear fit.} 
    \label{Fig.Difference_absorbance_3294_3187}
\end{figure}

\begin{figure}
    \centering
    \includegraphics[width=0.49\textwidth]{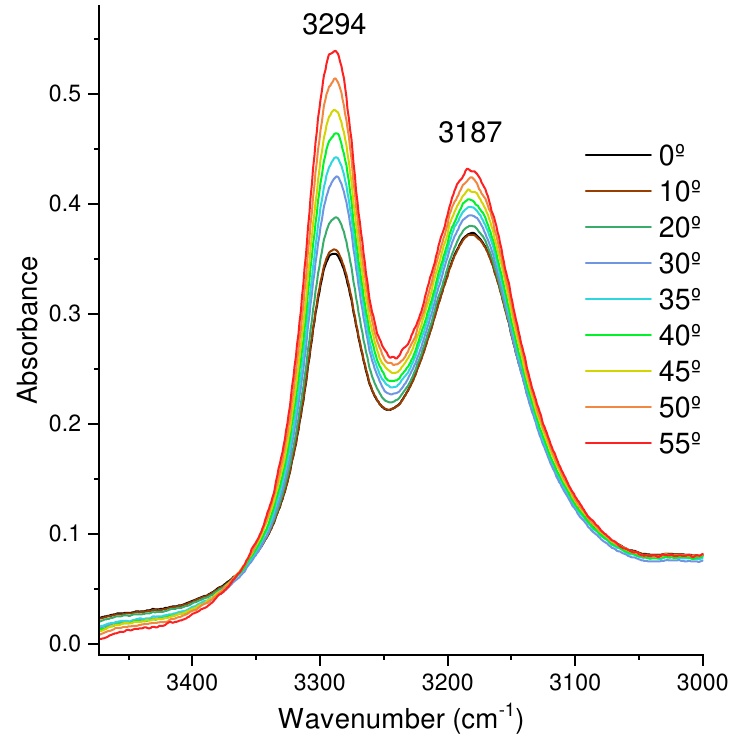}
    \includegraphics[width=0.49\textwidth]{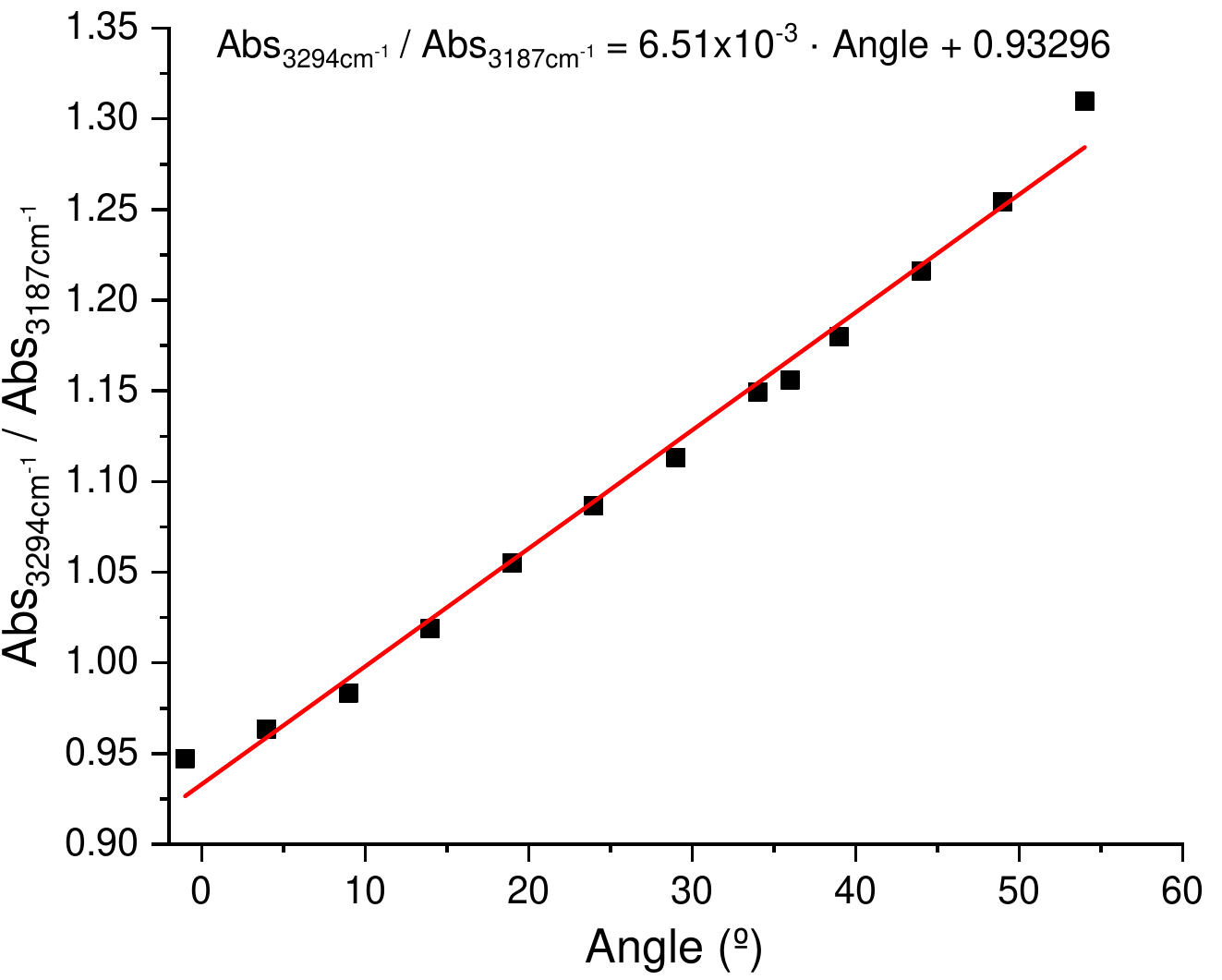}
    \caption{Top panel: IR spectra of a CH$_3$OH ice sample deposited at 125 K (Exp. \textbf{2}) measured at different incident angles. The ratio between the two components changes as a function of the incident angle. Bottom panel: ratio between the absorbance at the maximum of both peaks as a function of the incident angle with linear fit.}
    \label{Fig.IR_different_angles}
\end{figure}

Fig. \ref{Fig.IR_cristalizacion} presents the methanol ice IR spectra acquired at different warm-up temperatures ranging from deposition temperature at 30 K to 125 K. Above 115 K, there are no more observable changes in the IR spectrum until thermal desorption of the ice sample. The broad absorption band around 3238 cm$^{-1}$, attributed to the OH stretching of molecules in the solid phase, is split up into two bands at 113 K. This splitting becomes clear at 115 K in our experiments and has been reported in previous works \citep{Lucas2005, Merlin2012}. A discussion of the observed splitting is presented in Sect. \ref{sect:dft}.\\

The band strength that corresponds to the spectral range covering the O-H and C-H stretching modes of CH$_3$OH ice (Exp. \textbf{1}, Fig. \ref{Fig.Band_strength_density}), measured at different temperatures during warm-up, is shown in Fig. \ref{Fig.Band_strength_CH3OH}. The value of the band strength at the deposition temperature of 30 K, 2.3 $\times$ 10$^{-16}$ cm molecule$^{-1}$, which does not differ from the one at 10 K deposition, was calculated using Eq. \ref{eq.band_strength} with a methanol density of $\rho$ = 0.65 g cm$^{-3}$ and a refractive index of $ n$ = 1.265 adopted from \cite{Luna2018AA...617A.116L}. \cite{Hudgins1993} reported a value of 1.6 $\times$ 10$^{-16}$ cm molecule$^{-1}$ for methanol ice likely deposited at 10 K that is valid for the 3500 -- 2750 cm$^{-1}$ range; we note that these authors assumed a density of 1 g cm$^{-3}$ to estimate the band strength since this value was not known at the time. Our new values of the band strengths are 41 \% higher.\\

Using this band strength value as a starting point, the band strengths at higher temperatures during warm-up were obtained provided that the ice column density did not change as the result of the warm-up, which is a valid assumption prior to the onset of ice thermal desorption near 125 K. It was found that before this temperature was reached, there was an important variation in the band strength during crystallization of the ice. This sharp decrease of the band strength was accompanied by a sudden drop of the laser signal.\\

There are two important variations in the band strength that may be attributed to reestructuration of the amorphous structure, mainly between 60 and 100 K, and due to the crystallization above 110 K. During the first temperature range, there is an increase of the band strength with a maximum around 85 K, and a decrease of this value after 112 K. Laser signal variations at these temperatures were also detected.\\

Fig. \ref{Fig.IR_deposition_temperatures} compares the spectra of amorphous methanol deposited at 30 K (Exp. \textbf{1}); Exp. \textbf{1} warmed to 115 K; crystalline methanol deposited at 125 K (Exp. \textbf{2}); and crystalline methanol deposited at 135 K (Exp. \textbf{3}). The spectra display signatures specific of the different crystalline structures,  unveiling the thermal history, as can be seen in the shoulders appearing near 775 cm$^{-1}$, 973 cm$^{-1}$, 1141 cm$^{-1}$, and 2459 cm$^{-1}$, in addition to the change in profile of the main peaks. The appearance of an intense peak at 1299 cm$^{-1}$ is remarkable, as it suggest that this band is highly dependant on the deposition temperature and structure of the ice. This band is probably related to the CH$_3$ asym. bending located at 1330-1340 for low temperature experiments and 1310 cm$^{-1}$ for CH$_3$OH deposited at 125 K (see Table \ref{Table.frequencies}). Therefore, it is shifted about 30-40 cm$^{-1}$ for the different experiments.\\


Fig. \ref{Fig.Band_strength_density} is a representation of the band strength for three integration intervals as a function of temperature during warm-up. The spectral range from 3600 to 2700 cm$^{-1}$ covers the broad OH stretching band and the overlapping CH stretchings (blue triangles). Within the error bars this band strength is flat below ca. 100 K and slightly decreases at higher temperatures. However, when these band strengths, corresponding to the OH stretching (green circles) and CH stretchings (black squares), are represented individually it is observed that beyond 100 K the former decreases and the latter increases during warm-up. The density values for these warm-up temperatures (red stars) also remains constant around 0.65 g cm$^{-3}$ within the error bars up to about 100 K and gradually increases to attain 0.68 g cm$^{-3}$ at 130 K. It was not possible to estimate the density and band strength values at higher temperatures due to the onset of thermal desorption. \\

\begin{figure*}
    \centering
    \includegraphics[width=\textwidth]{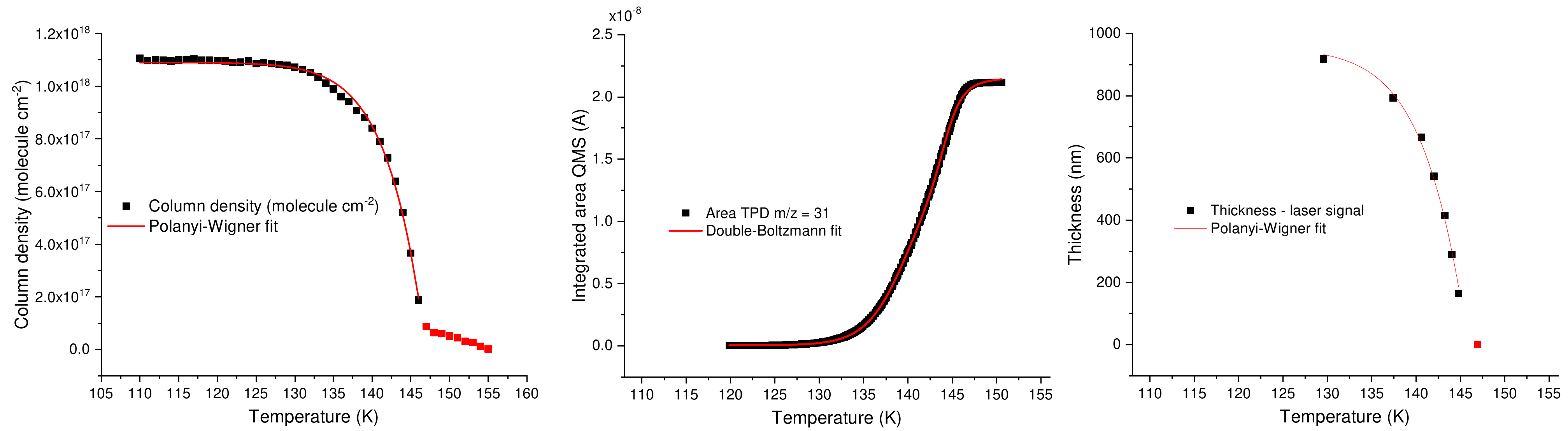}
    \caption{\small
    Thermal desorption of methanol ice. Left panel: Column density obtained from integration of the IR absorption band fitted with the Polanyi-Wigner equation. Middle panel: Integrated area below the QMS curve (for $\frac{m}{z} = 31$) with fit. Right panel: Estimated thickness of methanol ice during thermal desorption with fit. A constant value for methanol band strength ($2.22 \times 10^{-16}$ cm molec$^{-1}$ for the 3600 -- 2700 cm$^{-1}$ region) has been assumed during desorption. Functions used for adjusting data are the following:\\
    Left: $N \left( molecule \cdot cm^{-2}\right) = 1.1\times10^{18} - 1.6\times10^{31} \cdot e^{\frac{-4.4\times10^{3}}{T(K)}}$\\
    Middle: Area QMS $\left(10^{8} \cdot A\right) = 0.004 \;+\; 2.15 \cdot \left( \frac{0.66}{1 + e^{-\frac{T(K) - 139.78}{2.27}}}\right) \;+\; \left( \frac{0.33}{1 \;+\; e^{-\frac{T(K) - 144.04}{0.98}}}\right)$\\
    Right: Thickness (nm) =  953 $ - 1.81 \times10^{16} \cdot e^{\frac{-4.4\times10^{3}}{T(K)}}$\\
    Red squares in the figures represent data points which were not taken into account for the fits.\\
    }
    \label{Fig.fits}
\end{figure*}

The top panel of Fig. \ref{Fig.Difference_absorbance_3294_3187} displays the evolution of the methanol ice spectra grown at 125 K at different deposition thicknesses. The double peak corresponding to the OH stretching modes switches the predominance of the peak at lower wavenumbers for the peak at higher wavenumbers; the absorbances ratio of these peaks evolves linearly as the ice column density increases, see the bottom panel of Fig. \ref{Fig.Difference_absorbance_3294_3187}. The ice formed at this temperature should be predominantly crystalline. It is expected that this variation of the absorbance ratio is related to the oscillation of the corresponding modes at different orientations, which may be closer to the plane parallel or to the plane perpendicular to the substrate. This phenomenom was reported and explained in more detail for CO$_2$ ice deposition \citep{Escribano2013}.\\

More evidence for the different orientation of the two modes observed upon splitting of the OH stretching in crystalline methanol ice is presented in Fig. \ref{Fig.IR_different_angles}. The top panel displays IR spectra of a methanol ice sample deposited at 125 K (Exp. \textbf{2}) and collected at different incident angles of the IR beam with respect to the sample holder. The intensity ratio between the two bands changes as a function of the incident angle, $\theta_i$, where $\theta_i$ = 0$^{\circ}$ is the angle perpendicular to the sample substrate. 
For larger incident angles, the left band gains more intensity than the right one.
The bottom panel represents the ratio of these absorbances at the maximum of both peaks, as a function of the incident angle, and a linear fit. It is expected that the orientation of the OH stretchings for these bands follows (x,y,z) $\neq$ (0,0,0). In this angular range, where the left band gains more intensity, the plane parallel component of the transmittance increases while the perpendicular component decreases \citep{Hecht}, suggesting that the left band interacts more strongly with the plane parallel to the substrate. The detailed explanation of this effect will be presented in Sect. \ref{sect:dft}.\\

Figure \ref{Fig.fits} shows the sublimation of methanol ice as registered by the IR spectrometer (left panel), the QMS (middle panel), and the laser interference (right panel), see figure caption. The three plots show a maximum sublimation rate around 145 K. The decrease of the ice column density in the left panel is well fitted by the classical Polanyi-Wigner equation with the exception of an overfit between 130--140 K that is likely due to the transition from the $\alpha$ to $\beta$ crystalline structure. Those phases present different IR behavior as can be shown in Fig. \ref{Fig.IR_deposition_temperatures}. This diminution of the ice column density must be accompanied by the release of methanol molecules to the gas observed by the QMS, middle panel, which is well-fitted by a Double-Boltzmann function, see caption. The decrease in ice thickness is monitored by the laser interference displayed in the right panel, the Polanyi-Wigner equation also leads to a good fit in this case.\\  

The pre-exponential factor obtained from the Polanyi-Wigner fit of the column density is 1.6 $\times$ 10$^{31}$ molecule cm$^{-2}$. On the other hand, from the Polanyi-Wigner fit of the ice thickness derived from the laser signal, we have assumed a density of 0.68 g cm$^{-3}$ (see Fig. \ref{Fig.Band_strength_density}), obtaining a value of 2.3 $\times$ 10$^{31}$ molecule cm$^{-2}$. Both values are in good agreement with the one reported by \cite[][, see Fig. 7]{Luna2018AA...617A.116L}, while the desorption energy is lower because the maximum temperature of desorption, 145 K, is low compared to Luna et al. 2018, 163 K. This variation in the desorption temperature is due to the different experimental conditions, in particular high-vacuum setups tend to display higher desorption temperatures compared to ultra-high vacuum setups.

\subsection{Computational study} \label{sect:dft}
As shown in Fig. \ref{Fig.IR_cristalizacion}, the O-H stretching band around 3200 cm$^{-1}$ splits into two bands when the temperature increases. Understanding this splitting and the ones observed in other bands is particularly important because it marks the amorphous-crystalline phase transition and is indicative of the temperature of the ice.\\

\begin{figure*}
    \centering
    \includegraphics[width=0.75\textwidth]{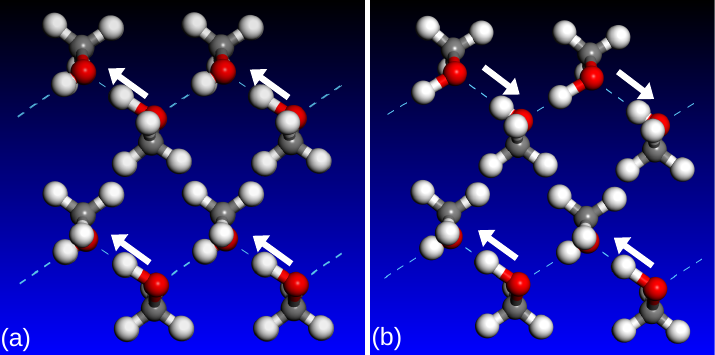}
    \caption{$\alpha$-phase of methanol ice. Dashed lines represent the OH$\cdots$O hydrogen bonds forming methanol chains. Arrows represent the motion of O-H stretching when neighbouring chains are in-phase (a) or in counter-phase (b).}
    \label{Fig.methanol_chain}
\end{figure*}

We complement our experimental results with DFT simulations of crystalline methanol. 
We use the low temperature $\alpha$-phase with four molecules in the unit cell, space group $P2_12_12_1$, cell parameters: a=4.6469\AA, b=4.9285\AA, c=9.0403\AA, $\alpha=\beta=\gamma=90^\circ$, as provided by \cite{Kirchner2008}. 
The methanol molecules are chained together by OH$\cdots$O hydrogen bonds, as seen in Fig. \ref{Fig.methanol_chain}.
The higher temperature $\beta$-phase belongs to the $Cmc2_1$ space group, forming similar chains but with a different orientation. 
The more distinguishable differences in the IR spectra of both phases are the apparent dilution of the OH-str. band splitting in the $\beta$-phase (\cite{Lucas2005}) and the different intensity ratio in the splitting of the band around 1500 cm$^{-1}$. Our spectra are more similar to the $\alpha$-phase as expected for the working temperature range in our experiments, we therefore focus this computational study on the $\alpha$-phase. \\

\begin{figure*}
    \centering
    \includegraphics[width=0.95\textwidth]{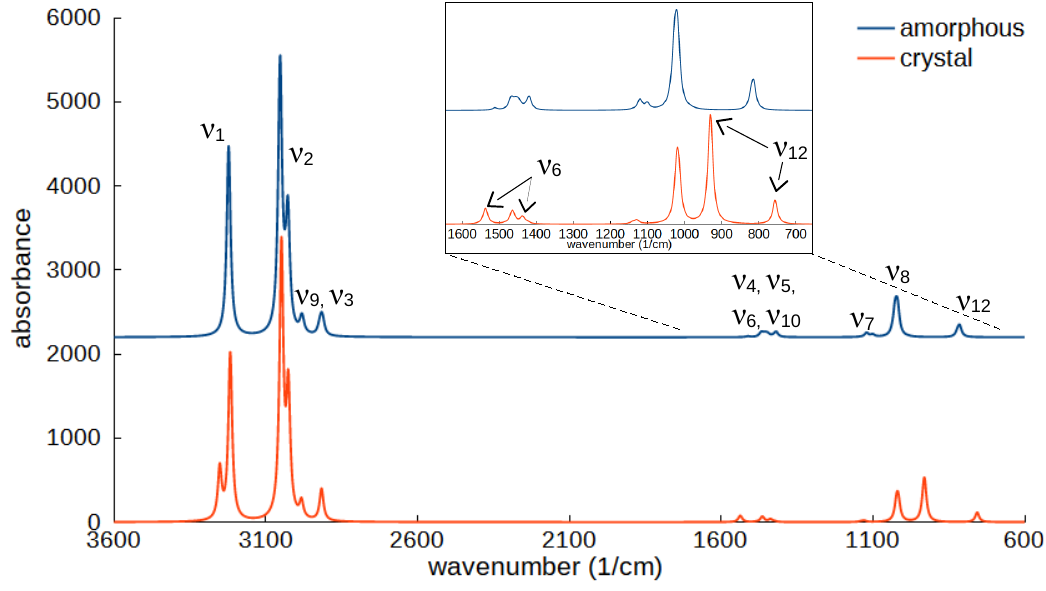}
    \caption{Simulated IR spectra for amorphous (top) and crystalline (bottom) methanol models. Curves are generated as Lorentzian curves at predicted vibrational modes with a FWHM of 15~cm$^{-1}$. Spectra are shifted for clarity.}
    \label{Fig.methanol_DFT_IR}
\end{figure*}

Using the CASTEP software \citep{CASTEP}, we performed geometry optimization of the initial structure until a well-defined minimum for the potential energy is found.
The fully optimized unit cell parameters are provided as Supplementary Information, along with bond lengths and angles.
We then perform a linear response calculation \citep{CASTEP_DFPT} and we generate a simulated IR spectrum (Fig. \ref{Fig.methanol_DFT_IR}). All the vibrational modes observed in the experiments are also found in our simulations with the exception of combination modes (Table \ref{Table.frequencies}). \\

The O-H stretching double band appears at 3206 and 3245~cm$^{-1}$ in our simulations. The slight divergence with experimental results is to be expected and can be attributed to the imperfections of the model.
In particular, contrary to the experiments, the CH-stretching modes are more intense than the OH-stretching mode in the modelled spectrum, and the CO stretching near 1026 cm$^{-1}$ in the experiment is much less intense in the simulation.
A similar model was built for the amorphous case by starting from the crystalline structure and using molecular dynamics with temperature annealing until all periodic order was lost. 
At this stage we apply an Andersen barostat \citep{Andersen1980} allowing the cell to expand until we reach a density of 0.65~g cm$^{-3}$, as reported by \cite{Luna2018AA...617A.116L} for amorphous methanol ice.
As expected, this case produces only a single, broader O-H stretching band centered at 3222~cm$^{-1}$.
In Fig. \ref{Fig.methanol_DFT_IR} we compare the resulting IR spectra for both cases. \\

We generated animations for the two vibrational modes in the crystalline O-H double band. A scheme of the two modes is shown in Fig. \ref{Fig.methanol_chain}, while the animation is available in the Supplementary Information. 
We find that all O-H bonds are coupled, oscillating in a synchronized motion within each methanol chain. 
This behavior is to be expected, as the hydrogen bonded chains act as one dimensional chain of harmonic oscillators. The distance between chains is approximately twice as the distance between molecules in the chain \citep{Kirchner2008}, so the interaction between chains is weaker.  \\

The difference between the two O-H stretching vibrational modes lies in the synchronization between chains. We can observe that the 3206~cm$^{-1}$ band corresponds to the coupling of neighboring methanol chains, where all the O-H bonds are oscillating in phase. Alternatively, the 3245~cm$^{-1}$ band corresponds to a counter-phase motion, where O-H bonds are still coupled within their chain but the neighboring chains have a 180$^\circ$ phase shift (see Fig. \ref{Fig.methanol_chain} and Suppl. Information for videos).
The two vibrational modes produce average dipoles in different directions, which would explain the angular dependence of the absorbance (Fig. \ref{Fig.IR_different_angles}). \\

In the amorphous case we do not observe a chain of molecules so there is no possible synchronized motion, resulting in a single band at 3222~cm$^{-1}$.
This is consistent with previous justifications for the double O-H stretching band \citep{Falk1961}, where the splitting was considered to be caused by the coupling of the motions of molecules in the unit cell.\\

Although it is less noticeable, we also observe a similar band doubling for the OH in-plane bending around $\sim$1500~cm$^{-1}$, as well as the OH out-of-plane bending at $\sim$700~cm$^{-1}$, the latter mode is referred to as torsion by some authors. As detailed in Fig. \ref{Fig.methanol_DFT_IR}, neither band doubling is observed for the amorphous case. The presence or absence of these splittings are in line with the experimental results reported here.  \\

In Table \ref{Table.frequencies} we summarize the vibrational frequencies for experiments and simulations. We include an amorphous ice deposited at 30 K corresponding to Exp. \textbf{1} in Fig. \ref{Fig.IR_deposition_temperatures}, the same ice after warm-up to 115 K, and finally an $\alpha$-crystalline ice deposited at 125 K (Exp. \textbf{2}) and $\beta$-crystalline at 135 K (Exp. \textbf{3}). 
We also include calculated frequencies from amorphous and crystalline simulations. In the amorphous case we indicate ranges of frequencies when it was not possible to obtain a single reproducible result.
The calculated frequencies are close to the experimental values and are helpful for assigning vibrational modes. Some bands are more accurate than others, which is to be expected from the applied model at this level of precision. 

\begin{table*} 
    \caption[]{Experimental and calculated vibrational frequencies (cm$^{-1}$) for solid methanol.}
    \label{Table.frequencies}
    \begin{tabular}{clcccccc}
            &         & \multicolumn{4}{c}{experiments} & \multicolumn{2}{c}{simulations} \\
	  Band    & band assignment         & T=30 K  & T=115 K  & T=125 K   & T=135 K  & amorphous  & crystalline \\
	  \hline
	\noalign{\smallskip}
 	$\nu_1$    & OH stretching        & 3233    & 3189, 3289   &3184, 3294   &3201, 3286 & 3222       & 3206, 3245 \\
	$\nu_2$    & CH$_3$ asym. stretch & 2982    & 2984             &2986   &2985       & 3027, 3053 & 3025, 3048 \\
	$\nu_9$    & CH$_3$ asym. stretch & 2954    & 2956             &2956   &2957       & 2978--2983 & 2982       \\
	$\nu_3$    & CH$_3$ sym. stretch  & 2828    & 2831             &2832   &2832       & 2912--2924 & 2916       \\
               & comb.                & 2606    & 2612             &2614   &2617       &            &            \\
               & comb.                & 2526    & 2535             &2535   &2564, 2535 &            &            \\
               &                      &         & 1509             &1510   &1510       &            &            \\
    $\nu_4$    & CH$_3$ asym. bending & 1458    & 1472             &1472   &1472       & 1454       & 1483       \\
    $\nu_5$    & CH$_3$ sym. bending  & 1445    & 1458             &1456   &1372, 1405 & 1447       & 1464       \\ 
    $\nu_{10}$ & CH$_3$ asym. bending & 1332    & 1338             &1310   &1299       & 1419--1425 & 1461       \\ 
	$\nu_6$    & OH bending           &         &                  &   &           & 1419--1425 & 1437, 1536 \\ 
	$\nu_7$    & CH$_3$ rocking       & 1129    & 1143             &1142   &1141       & 1101--1121 & 1128, 1140 \\ 
	$\nu_8$    & CO stretching        & 1027    & 1025             &1021, 1025, 1032   &1021, 1033 & 1018--1027 & 1021       \\ 
    $\nu_{12}$ & CH$_3$OH torsion     & 727     & 678, 771         &677, 766   &674, 757   & 814--821   & 756, 929   \\ 
 	\noalign{\smallskip}
	\hline
    \end{tabular}\\
\end{table*} 

\section{Conclusions} \label{Conclusions}

The physical properties of methanol ice with density and IR spectroscopy as the primary goals, were studied at different deposition temperatures and during the warm-up. The IR band strengths are provided. The transition from amorphous to crystalline methanol ice was also explored. The following conclusions are drawn:

\begin{itemize}
    \item The volumetric density and the various IR band strengths of methanol tend to remain rather constant during warm-up of the ice from 10 K until nearly 100 K, see Fig. \ref{Fig.Band_strength_density}. Beyond 100 K there is a significant increase of the density, and the IR band strength values also vary at these high temperatures until thermal desorption of the ice takes place. These changes are attributed to the crystallization process of amorphous methanol ice during warm-up. In comparison, the methanol ice density of samples deposited at different temperatures \citep{Luna2018} experienced larger variations at temperatures below $\sim$100 K, while the opposite effect is observed above this temperature. The likely explanation is that a more ordered structure is attained when the ice is grown at the higher temperatures, what is in concordance with the IR spectra shown in Fig. \ref{Fig.IR_deposition_temperatures}. If, instead, methanol ice is grown near 10 K, crystallization above 100 K causes an increase of the ice density and more prominent spectroscopic changes.\\
    
    \item The observed splitting into two bands of the O-H stretching band around 3200 cm$^{-1}$ is indicative of the amorphous-crystalline phase transition during ice warm-up. The crystalline $\alpha$ and $\beta$ phases correspond to molecules chained by OH$\cdots$O bonds with different orientations.\\ 
    
    \item The CO stretching band near 1025 cm$^{-1}$ does not change its profile during warm-up. This band, however, displays a profile that depends on the deposition temperature, see Fig. \ref{Fig.IR_deposition_temperatures}. The astronomical observations of this band are discussed in Sect. \ref{astro_implications}.\\
    
    \item The vibrational assignments of the IR bands was revisited using a computational model. It was found that the splitting of the O-H stretching band is caused by synchronization between chains: while the 3206 cm$^{-1}$ band is attributed to coupling of neighboring methanol chains with O-H bonds oscillating in phase, the 3245 cm$^{-1}$ band corresponds to a counter-phase motion with O-H bonds still coupled within their chain but the neighboring chains have a 180$^\circ$ phase shift. The model enabled us to generate an animation for better visualizing the different modes (Suppl. Information). 
    Our computational results are consistent with \cite{Falk1961}, who explained the band splitting as the two possible couplings, in and out of phase, and proved the splitting disappears when using isotopes that cannot synchronize their oscillation.\\
    
    \item A less noticeable band splitting of the OH in-plane bending around 1500 cm$^{-1}$ and the OH out-of-plane bending around 700 cm$^{-1}$ was observed. 
\end{itemize}

\section{Astrophysical implications} \label{astro_implications}

Methanol formation is believed to occur in regions protected from radiation deep inside dense
clouds or protostellar envelopes, e.g. \cite{Boogert2022} and references therein. The identification of methanol as one of the ice mantle components toward protostars and specific lines of sight toward dense clouds, is mainly supported by the presence of absorption bands centered at 3.54, 3.84, and 3.94 $\mu$m, respectively 2825, 2604, and 2538 cm$^{-1}$ that correspond to the $\nu_3$ CH$_3$ symmetric stretch and combination modes \citep{Grim1991, Chiar1995, Dartois1999, Dartois2003}. \cite{Dartois2003} provides a good fit of the combination modes in this range with amorphous methanol deposited by vapor deposition at 10 K in the laboratory. With a sufficiently high spectral resolution, it might be possible to elucidate the formation temperature of methanol ice present in ice mantles by fitting the band profiles with the laboratory spectra, see the middle panel of Fig. \ref{Fig.IR_deposition_temperatures}.    
The OH-stretching modes of methanol are sensitive to the degree of crystallization in the ice, but these bands overlap with the OH-stretching of water in astrophysical ices, thus hindering the comparison of observations with the reported experimental methanol spectra. If observed, the band splittings of the OH in-plane bending around 1445 cm$^{-1}$ (6.92 $\mu$m) and the OH out-of-plane bending around 727 cm$^{-1}$ (13.75 $\mu$m), would also be indicative of crystallized methanol ice.\\

The CO stretching band near 1025 cm$^{-1}$ deserves special attention. This band profile is consistent with a significant fraction of the CH$_3$OH being in a relatively pure or CO-rich
phase (e.g. \cite{Pontoppidan2003, Bottinelli2010}). We report that the profile of this band is sensitive to the deposition temperature of pure methanol ice, its splitting becomes most evident when deposited at high temperatures, above 130 K in our experiments, likely corresponding to a $\beta$-crystalline structure (see Fig. \ref{Fig.IR_deposition_temperatures}). Two bands appear at 1033 cm$^{-1}$ (9.68 $\mu$m) and 1021 cm$^{-1}$ (9.78 $\mu$m), respectively. Unfortunately, this two-component profile could not be recognized at the spectral resolution and signal/noise of the reported Spitzer data \citep{Bottinelli2010}. New observations of this band with JWST at higher resolution would be capable of resolving a double-peaked profile of this band. The exact band position and the profile of this absorption band could serve to identify the presence of pure/mixed methanol and an amorphous/crystalline structure.\\      

In addition to the assessments of a pure/mixed and amorphous/crystalline structure of methanol in the observed ice mantles, the information extracted from the experiments is useful to constrain other methanol ice properties such as density and IR band strength. It was found that the density and IR band strengths of methanol remain rather constant during warm-up of the ice from 10 K until nearly 100 K, see Fig. \ref{Fig.Band_strength_density}. Beyond 100 K there is a significant increase of the density, and the IR band strength values also vary at these high temperatures until thermal desorption of the ice takes place. The measured values of the IR band strengths at different temperatures serve to provide better estimates of the methanol column densities based on the observed IR bands of methanol. Compared to previous works \citep{Hudgins1993} that simply assumed a density of 1 g cm$^{-3}$ for methanol ice, the significantly lower density values measured for methanol ice, \cite{Luna2018} and this paper, lead to values of the IR band strength for various methanol bands about 40 \% higher, and thus provide an estimation of the methanol ice column density a factor of 40 \% lower. For instance, for the lowest methanol deposition temperatures we obtain a band strength of 2.3 $\times$ 10$^{-16}$ cm molecule$^{-1}$ for the integration range between 3500--2750 cm$^{-1}$.\\

\section*{Acknowledgements}
\label{acknowledgements}
This research has been funded by projects PID2020-118974GB-C21 and 
PID2020-118974GB-C22 by the Spanish Ministry of Science and Innovation.
B.E. acknowledges support by grant PTA2020-018247-I by the Spanish Ministry of Science and Innovation/State Agency of Research MCIN/AEI. We thank our colleagues Pedro G\'omez at UCM and Vicente Tim\'on at IEM for their helpful advice and discussions.


\section*{Data Availability}
The data underlying this article cannot be shared publicly. 


\bibliographystyle{mnras}
\bibliography{bibliography} 

\bsp	
\label{lastpage}
\end{document}